# Enabling *P*-type Conduction in Bilayer WS$_2$ with NbP Topological Semimetal Contacts

Lauren Hoang[1,#], Asir Intisar Khan[1,#], Robert K.A. Bennett[1], Hyun-mi Kim[2], Zhepeng Zhang[3], Marisa Hocking[3], Ae Rim Choi[4], Il-Kwon Oh[4], Andrew J. Mannix[3,5*], and Eric Pop[1,3,6,7*]

[1]*Department of Electrical Engineering, Stanford University, Stanford, CA 94305, USA*
[2]*Korea Electronics Technology Institute, Seongnam-si 13509, Republic of Korea*
[3]*Department of Materials Science & Engineering, Stanford University, Stanford, CA 94305, USA*
[4]*Department of Intelligence Semiconductor Engineering, Ajou University, Suwon 16499, Republic of Korea*
[5]*Stanford Institute for Materials & Energy Sciences, SLAC National Accelerator Lab., Menlo Park, CA 94025, USA*
[6]*Department of Applied Physics, Stanford University, Stanford, CA 94305, USA*
[7]*Precourt Institute for Energy, Stanford University, Stanford, CA 94305, USA*

[#]*These authors contributed equally to this work.*
[*]*E-mail: ajmannix@stanford.edu, epop@stanford.edu*

**Two-dimensional (2D) semiconductors are promising for low-power complementary metal oxide semiconductor (CMOS) electronics, which require ultrathin *n*- and *p*-type transistor channels. Among 2D semiconductors, WS$_2$ is expected to have good conduction for both electrons and holes, but *p*-type WS$_2$ transistors have been difficult to realize due to the relatively 'deep' valence band and the presence of mid-gap states with conventional metal contacts. Here, we report topological semimetal NbP as *p*-type electrical contacts to bilayer WS$_2$ with up to 5.8 µA/µm hole current at room temperature; this is the highest to date for sub-2 nm thin WS$_2$ and > 50× larger than with metals like Ni or Pd. The *p*-type conduction is enabled by the simultaneously high work function and low density of states of the NbP, which reduce Fermi level pinning. These contacts are sputter-deposited at room temperature, an approach compatible with CMOS fabrication, a step towards enabling ultrathin WS$_2$ semiconductors in future nanoelectronics.**

Computing in the 21$^{st}$ century must process increasingly complex data loads, with recent large language models estimated to use as much energy as a small city every day[1]. From the hardware side, reducing energy use could be addressed by three-dimensional (3D) integration of logic and memory with dense interconnects, an approach that could yield 100× or higher energy efficiency per function[2]. Dense 3D integration calls for materials stacked at sufficiently low temperatures to avoid damaging existing (memory or logic) layers[3,4], and low-power operation requires complementary metal oxide semiconductor (CMOS) technology, with both *n*- and *p*-type transistors[5,6]. Energy-efficient, 3D electronics could benefit from adopting two-dimensional (2D) semiconductors, in nanosheet transistors[7] or in back-end-of-line (BEOL) logic and memory layers, due to their favorable mobility in sub-2 nm thin films[8,9] and their compatibility with lower temperature fabrication[10,11] than crystalline silicon.

Among 2D semiconductors, tungsten disulfide ($WS_2$) is one of the few expected to have simultaneously good electron and hole mobility[9], allowing it to serve as a single material for both *n*- and *p*-type future transistors, unlike $MoS_2$ and $WSe_2$. $WS_2$ also has one of the larger band gaps among 2D semiconductors ($> 2.2$ eV for bi- or monolayers)[12], an advantage for low power operation. However, its valence band is 'deeper', with respect to the vacuum level, than the Fermi level of most common metals with work function up to 5.6 eV; this is a key reason why *p*-type contacts have been difficult to achieve with $WS_2$, compared to $WSe_2$ or other 2D materials having 'shallower' valence bands. Conventional metals may also cause metal- and defect-induced gap states (MIGS and DIGS) that pin the Fermi level unfavorably (**Fig. 1a**), and although this effect could be reduced with Bi or Sb for *n*-type devices[13], the low work function of such semimetals limits their use for *p*-type contacts.

Here we achieve record *p*-type current in bilayer $WS_2$ (~1.3 nm thin) with NbP semimetal contacts. These semimetals are expected to have large work function based on recent density functional theory (DFT) calculations[14], and their low density of states near the Fermi level could limit Fermi level pinning, enabling improved hole injection (**Fig. 1b**). NbP is also a Weyl semimetal, previously found to show signatures of topological protection in single crystal[15] and ultrathin nanocrystalline films[16] similar to $Bi_2Se_3$[17]. When interfaced with bilayer $WS_2$, such topological semimetals could display suppressed MIGS, and their large work function better aligns the Fermi level with the bilayer $WS_2$ valence band (**Fig. 1b,c**, **Supplementary Fig. S1**), both desired features for *p*-type conduction.

Importantly, our NbP contacts were sputter-deposited directly onto $WS_2$ at room temperature, a process compatible with 3D integration and BEOL fabrication[3] (see **Methods** for details). However, NbP contacts sputtered on *monolayer* $WS_2$ were found to cause some defects (see **Supplementary Fig. S2 and S3a**); thus, we focus primarily on *bilayer* $WS_2$ in this study, which has higher immunity to process damage (**Supplementary Fig. S3b**) and a higher valence band[12] than monolayer $WS_2$.

To examine the interface of NbP contacts with bilayer $WS_2$, we performed high-resolution scanning transmission electron microscopy (STEM) across such a stack on $SiO_2$/Si (**Fig. 2a-c**), similar to the transistor contacts employed further below. Here, the bilayer $WS_2$ films were directly grown by chemical vapor deposition (CVD)[18] onto the $SiO_2$/Si substrate. STEM images show the conformal deposition of ~3.5 nm sputtered NbP onto bilayer $WS_2$ without obvious defects (**Fig. 2a**).

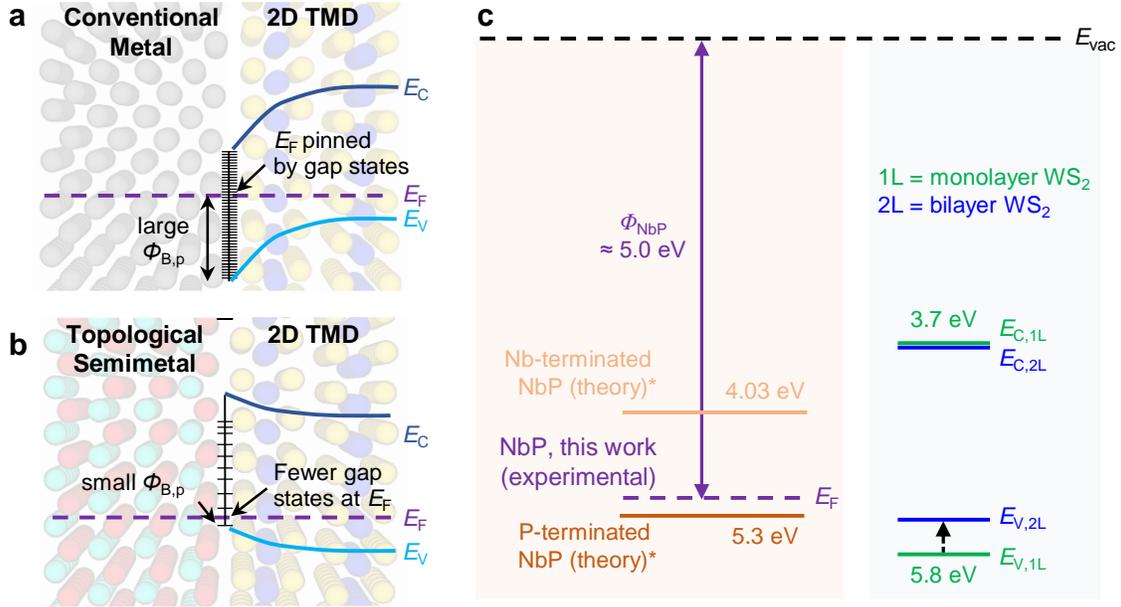

**Figure 1 | Metal and topological semimetal NbP interfaces with WS$_2$. a,** Conventional metal, and **b,** topological semimetal interfaced with a 2D semiconductor (here, WS$_2$). The conventional metal tends to cause significant Fermi level ($E_F$) pinning in the band gap, leading to large Schottky barrier height for hole conduction ($\phi_{B,p} = E_F - E_V$). In contrast, topological semimetals with large work functions, such as NbP (and TaP), are expected to have less pinning and lower hole barrier height. $E_C$ and $E_V$ denote the conduction and valence band edges, respectively. **c,** Schematic diagram of NbP–WS$_2$ band alignments. The valence band of bilayer (2L) WS$_2$ is higher than that of monolayer (1L), and better aligned with our experimental work function of NbP (see **Supplementary Fig. S1**). The expected work function of NbP (marked by *) and the WS$_2$ bands in the figure are based on Refs. 12,14.

The elemental energy dispersive spectra (EDS) in **Fig. 2d-h** confirm uniform distribution of Nb and P in our NbP film on bilayer WS$_2$. Atomic force microscope (AFM) mapping also reveals low surface roughness and continuous NbP both on SiO$_2$ (**Supplementary Fig. S4a**) and WS$_2$ (**Fig. 2i** and **Supplementary Fig. S4b**). Previous work has found that many metals deposited on 2D semiconductor surfaces tend to aggregate and form clusters[19,20]. However, our NbP deposited on WS$_2$ has no measurable change in surface roughness compared to when sputtered on SiO$_2$. Both top-down AFM and cross-sectional TEM imaging suggest intimate NbP-WS$_2$ contacts, without agglomeration. We also used Raman spectroscopy before and after NbP deposition, finding that the E$_g$ and A$_{1g}$ peaks of bilayer WS$_2$ (**Fig. 2j**) are essentially unaffected by the deposition of NbP (of either ~2 nm or ~3.5 nm thickness), unlike for monolayer WS$_2$ (**Supplementary Fig. S3a**). However, the formation of some defects is not unexpected[20,21] after direct sputtering of NbP on WS$_2$, as suggested by the increased LA(M) peak intensity[22] (**Supplementary Fig. S3b-d**), especially as the number of WS$_2$ layers is decreased.

We fabricated bilayer WS$_2$ transistors with NbP and TaP contacts (**Fig. 3a,b**), capped by Pd/Au, as well as control devices with only Pd/Au contacts. Devices have channel lengths $L$ between 100 nm and 1 μm, channel widths of 1 or 2 μm, and are back-gated by the heavily doped Si substrate through

100 nm $SiO_2$. A top-down scanning electron microscope (SEM) image of a $WS_2$ bilayer 'strip' with multiple device channels is shown in **Fig. 3b**, and additional fabrication details are given in **Methods**.

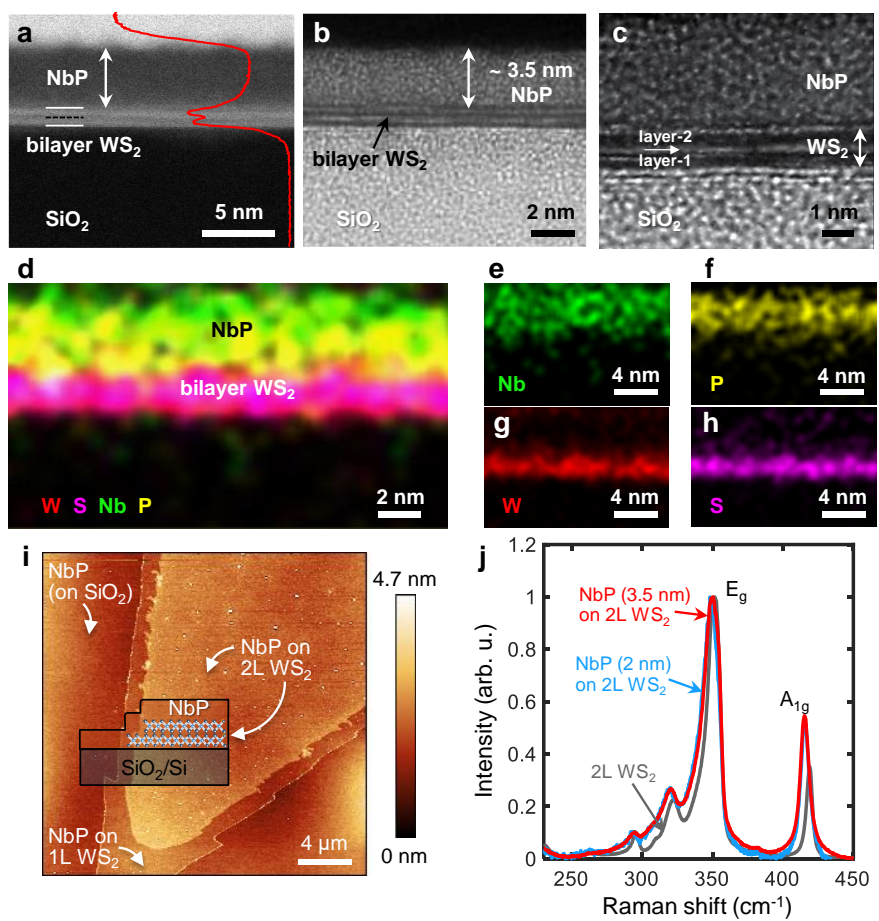

**Figure 2 | Characterization of the NbP/bilayer $WS_2$ interface. a,** High-angle annular dark field imaging (HAADF) scanning transmission electron microscopy (STEM) cross-section of NbP sputtered onto bilayer $WS_2$ on an $SiO_2$/Si substrate. STEM shows ~3.5 nm NbP film conformally deposited on the ~1.3 nm thin bilayer $WS_2$ (horizontal lines mark the edges of the layers and the van der Waals gap between them). **b,** Higher-resolution, and **c,** zoomed-in STEM images of the same sample. **d,** Energy dispersive spectroscopy (EDS) showing the elemental distribution of Nb, P, W, and S (combined), and of **e,** Nb, **f,** P, **g,** W, and **h,** S atoms (separately). The EDS measurement scan time was short to minimize electron-beam damage of the NbP/$WS_2$ stack and interface. This led to a lower-resolution EDS as a trade-off. **i,** Surface mapping with atomic force microscopy of ~3.5 nm NbP on bilayer $WS_2$ and nearby $SiO_2$ substrate, showing similar low surface roughness on both regions (~0.16 nm) (see **Supplementary Fig. 4**). Inset shows a cross-section schematic of the NbP covering steps of monolayer (1L) and bilayer (2L) $WS_2$. **j,** Raman spectra (532 nm laser, at room temperature) of bilayer $WS_2$ before (gray) and after (blue and red) being covered by NbP of two different thicknesses (~2 nm and ~3.5 nm), all showing similar $E_g$ and $A_{1g}$ peaks (normalized to the $E_g$ peak intensity). The $E_g$ and $A_{1g}$ peaks correspond to in-plane and out-of-plane atomic vibrations, respectively.

**Fig. 3c** displays measured drain current vs. gate voltage ($I_D$ vs $V_{GS}$) of bilayer $WS_2$ transistors, revealing >100× larger $|I_{D,max}|$ for our NbP-contact devices than our control Pd-contact devices. Similarly, our bilayer $WS_2$ devices with TaP contacts (another topological semimetal candidate) also show enhanced p-type current compared to control Pd contacts (**Supplementary Fig. S5a**). When probing the positive $V_{GS}$ range (**Supplementary Fig. S6**), Pd-contacted bilayer WS$_2$ devices also show stronger *n*-type behavior, similar to previous reports on ultrathin WS$_2$ using traditional metal contacts[22,23].

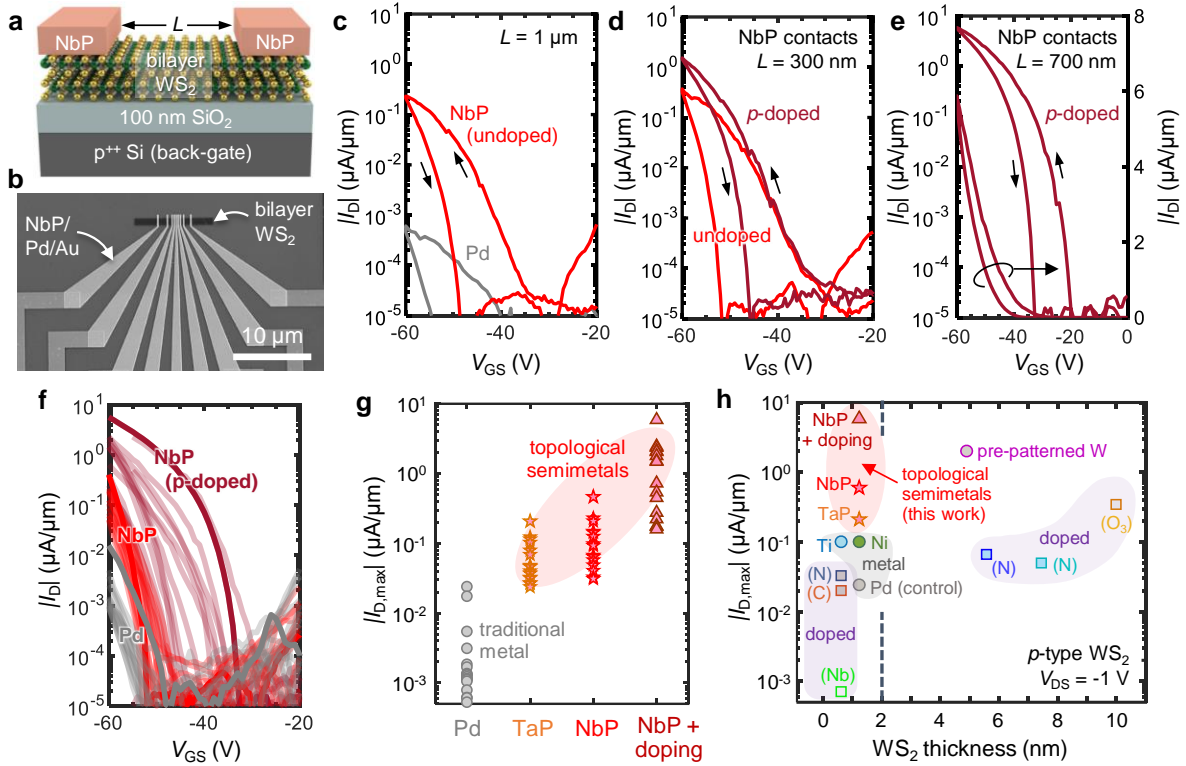

**Figure 3 | Bilayer WS$_2$ transistors with NbP and TaP semimetal contacts. a,** Schematic of a back-gated bilayer WS$_2$ transistor, not to scale. **b,** Top-view SEM image of fabricated devices with NbP contacts capped by Pd/Au. **c,** Measured $I_D$ vs. $V_{GS}$ at $V_{DS}$ = -1 V for bilayer WS$_2$ transistors with NbP contacts (red) and Pd contacts (gray), at channel length $L$ = 1 µm. Small arrows mark forward and backward sweeps[24], revealing similar hysteresis with both contact types. **d,** $I_D$ vs. $V_{GS}$ at $V_{DS}$ = -1 V for a different bilayer WS$_2$ device ($L$ = 300 nm) with NbP contacts before (red) and after (dark red) additional *p*-type doping, achieved by soaking the device in chloroform. **e,** $I_D$ vs. $V_{GS}$ at $V_{DS}$ = -1 V showing maximum hole current of 5.8 µA/µm in another bilayer WS$_2$ device ($L$ = 700 nm) with NbP semimetal contacts and additional *p*-doping. $I_D$ is shown on both linear and log scale axes. **f,** Measured $I_D$ vs. $V_{GS}$ at $V_{DS}$ = -1 V for 15 bilayer devices (each) with Pd contacts (gray), NbP contacts (red), and NbP contacts with additional doping (dark red). Darker color lines represent the highest-current device of each kind. **g,** Maximum *p*-type current $|I_{D,max}|$ achieved at $V_{DS}$ = -1 V of devices with various contacts, including Pd, TaP (another topological semimetal), and NbP (before and after additional doping); 15 devices (symbols) are shown for each type. **h,** Benchmarking maximum *p*-type current $|I_{D,max}|$ vs. WS$_2$ (channel) thickness at $V_{DS}$ = -1 V with various contact materials and doping strategies[25–32], including NbP and TaP semimetals, and control Pd metal from this work. (Text labels in parenthesis next to the symbols represent the doping type used.) Our results with NbP semimetal contacts enable the highest hole currents to date (up to 5.8 µA/µm) achieved in bilayer WS$_2$, >50× larger than previous reports with sub-2 nm channel thickness. All devices in panels **c-g** are between 1 and 2 µm wide, and the $I_D$ reported is normalized (µA/µm) by the channel widths[24].

To further increase the *p*-type current of our NbP-contacted bilayer WS$_2$ devices we applied chloroform doping[33,34] (see **Methods**). With this, we measured $|I_{D,max}|$ up to 5.8 µA/µm at $V_{DS}$ = -1 V, nearly an order of magnitude higher than before doping, as shown in **Fig. 3d-f** and **Supplementary Fig. S7a,b**. The chloroform-doped devices also show a positive shift of threshold voltage, another signature of $p$-type doping. These doped WS$_2$ devices are surprisingly stable over time (up to 16 days), despite being simply stored in a nitrogen environment, uncapped (**Supplementary Fig. S7c,d**).

We summarize the maximum $p$-type current achieved from various bilayer WS$_2$ devices using different contacts in **Fig. 3g** (15 devices for each contact type). Despite the device-to-device variability, much of it inherent to academic nanofabrication, we note a clear trend of enhanced $p$-type conduction with our semimetal contacts (TaP and NbP) vs. conventional Pd contacts. With NbP contacts and doping, the $p$-type current is increased by two orders of magnitude vs. Pd contacts to bilayer WS$_2$.

We also note that the observed hysteresis remains a challenge in $p$-type TMDs on SiO$_2$[35] (compared to $n$-type MoS$_2$), and is often not reported. The hysteresis likely occurs due to the alignment of the TMD valence band with trap states in the SiO$_2$[36], due to interface adsorbates remaining from the TMD layer transfer process[37], and due to the imperfect nature of the as-grown materials today[38]. These drawbacks are not fundamental, and are expected to be resolved as the quality of dielectrics, TMDs, and their interfaces improves. Nevertheless, fair device-to-device comparisons can be made when sufficient devices are measured and only the contacts are changed, as in **Fig. 3f,g**.

Using NbP and TaP semimetal contacts, we achieved the highest $p$-type currents to date in sub-2 nm thin WS$_2$ (**Fig. 3h**). Including doping, our devices reached ~5.8 µA/µm at $V_{DS}$ = -1 V, over 50× higher than previous reports in this WS$_2$ thickness regime. In contrast, traditional metal contacts like Ti[27], Ni[28], or our control Pd show much suppressed $p$-type behavior on ultrathin WS$_2$. Thicker WS$_2$ (e.g., ~5 nm) grown from pre-patterned tungsten contacts does show $p$-type conduction[30]; however, this thickness offers no electrostatic benefits over silicon in nanoscale transistors. **Fig. 3h** also displays additional data with traditional metal contacts to WS$_2$, where relatively low $p$-type currents were obtained after substitutional doping (using Nb[29], C[26], N[25,31]) or charge-transfer doping (e.g., WO$_x$[32]).

We note that the highest measured $|I_{D,max}|$ in our devices is limited by the largest $|V_{GS}|$ we can apply without compromising the back-gate dielectric integrity. In addition, the valence band of bilayer WS$_2$ is sufficiently 'deep' so as to render the threshold voltage quite negative here. This means that, even at our highest $|V_{GS}|$ = 60 V, not all our devices are fully turned on, especially not the Pd-contacted ones (**Fig. 3f**). Nevertheless, we can estimate an upper bound of ~ 80 kΩ·µm for the $p$-type contact resistance of NbP to bilayer WS$_2$, as half the total resistance of the device with the highest current. Although better $p$-type contacts have been achieved with WSe$_2$ (which has a 'shallower' valence band)[14], our estimated contact resistance to WS$_2$ is presently ~50× lower than any previous results with this 2D semiconductor at sub-2 nm thickness (see **Supplementary Fig. S8**).

7To gain a deeper understanding of what enables the *p*-type behavior in NbP-contacted bilayer WS$_2$, we performed DFT simulations using Quantum ESPRESSO[39]. We perform fixed-cell relaxations of the contact schematic shown in **Fig. 4a**, then extract the DOS and the hole Schottky barrier height $\phi_{B,p} = E_F - E_V$, as labeled on **Fig. 4b,c**. We estimate the density of metal-induced gap states (MIGS) by integrating the DOS across the band gap. The locations of WS$_2$ band edges are approximate, due to the presence of MIGS, and further simulation details are given in the **Methods** section.

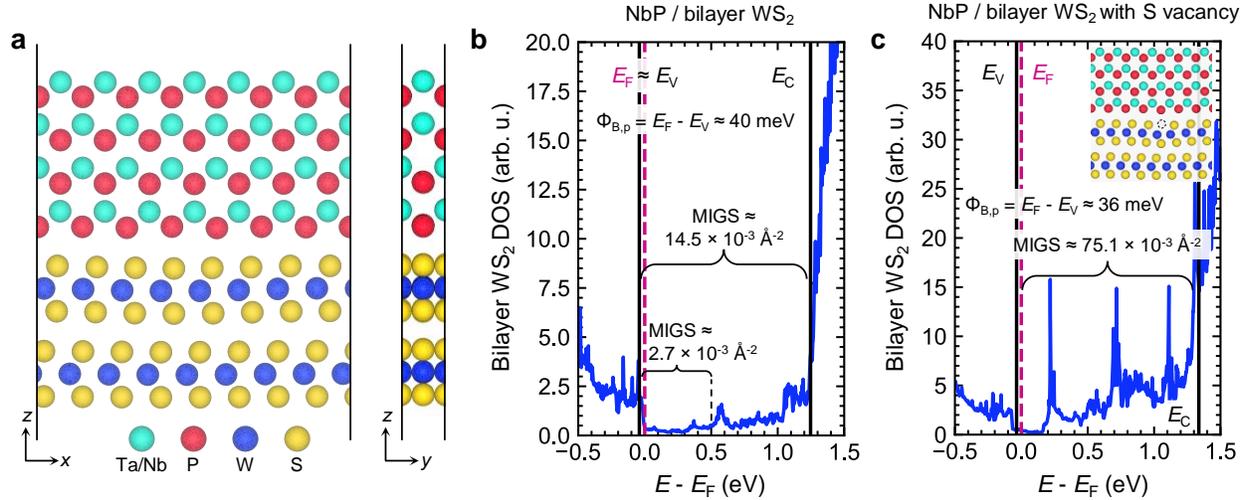

**Figure 4 | Density Functional Theory (DFT) Simulations. a**, Cross-section schematics of the DFT supercell for the semimetal contact (NbP or TaP) with bilayer WS$_2$. (Left) *x-z* cut and (right) *y-z* cut. **b**, Projected density of states (pDOS) contributions from bilayer WS$_2$ to the overall DOS of the NbP contact with bilayer WS$_2$. The hole Schottky barrier height is ≈ 0 meV. **c**, pDOS for the same contact but with one S vacancy (defect density of ~1.4 × 10$^{14}$ cm$^{-2}$), displaying small hole Schottky barrier height, ≈ 36 meV. The inset displays the supercell with the sulfur vacancy marked by a black dotted line. In panels **b-c**, the Fermi energy $E_F$ is set to zero and marked with a dashed purple line; the valence band maximum $E_V$ and conduction band minimum $E_C$ are marked with solid black lines. Locations of band extrema are determined from projected band structures (see **Supplementary Section II**). MIGS densities are calculated by integrating the bilayer WS$_2$ pDOS across the denoted energy ranges.

In **Fig. 4b**, we estimate a small Schottky barrier ($\phi_{B,p} \approx 40$ meV) for the NbP contact with pristine bilayer WS$_2$, at first without defects. This behavior is comparable to the expected band alignment for an ideal scenario, without Fermi level pinning ($\phi_{B,p} \approx -100$ meV as detailed in **Supplementary Section I**). These results also suggest weak Fermi level pinning at the NbP/bilayer WS$_2$ interface because of the low MIGS densities formed, between ~2.7 × 10$^{-3}$ Å$^{-2}$ and 14.5 × 10$^{-3}$ Å$^{-2}$, depending on the energy range of the DOS integral (the lower value being within 0.5 eV of $E_V$, where states are more likely to play a role in $E_F$ pinning). We also simulate the bilayer WS$_2$ contact with TaP and find a hole barrier height of 14 meV, unlike the ideal pinning-free scenario of -200 meV (**Supplementary Fig. S9a**); however, experimentally we find TaP does not provide contacts as good as NbP to bilayer WS$_2$ (**Fig.**

**3g,h**), ostensibly due to greater damage during deposition. In contrast to these semimetals, our simulations for a conventional metal contact (W) on bilayer WS$_2$ reveal a large hole Schottky barrier ≈ 800 meV and large MIGS density (~176.3 × 10$^{-3}$ Å$^{-2}$), as shown in **Supplementary Fig. S9b,c**. The estimated MIGS densities at our NbP/bilayer WS$_2$ contacts are also smaller than those of various conventional metal/2D semiconductor contacts from the literature[14,40].

We recall that the direct deposition of semimetals or conventional metals[20,21] onto 2D semiconductors tends to create defects. To understand this, we performed additional DFT simulations (**Fig. 4c**) after introducing a sulfur (S) vacancy in the top WS$_2$ layer, corresponding to a high defect density of ~1.4 × 10$^{14}$ cm$^{-2}$. Even in this case, we estimate a nearly unchanged $\phi_{B,p}$ ≈ 36 meV at our NbP contacts with bilayer WS$_2$, smaller than most metal-WS$_2$ systems[14,40]. Together, the favorable band alignment between NbP and bilayer WS$_2$, with sufficiently small MIGS densities (which prevent strong Fermi level pinning) support the favorable *p*-type conduction observed in our experiments with NbP contacts. Our calculations suggest a larger, non-negligible hole Schottky barrier for NbP with *monolayer* WS$_2$ ($\phi_{B,p}$ ≈ 550 meV in **Supplementary Fig. S10**). This appears to occur due to the larger band gap and electron affinity (i.e., 'deeper' valence band) of monolayer WS$_2$ compared to bilayer WS$_2$.

In summary, we have introduced the topological semimetal NbP for improved *p*-type contacts to thin, bilayer WS$_2$. Together with doping, such NbP-contacted WS$_2$ transistors reach 5.8 µA/µm hole current (at $V_{DS}$ = -1 V), over 50× larger than previous *p*-type WS$_2$ devices with sub-2 nm thickness. The improved *p*-type performance is enabled by the simultaneously large work function and low density of states of the semimetal contacts, leading to small hole barrier heights. Notably, these contacts are deposited by sputtering at room temperature, a back-end-of-line compatible approach. These results provide physical insight into contact formation to the valence band of 2D semiconductors and demonstrate a possible route towards their incorporation into future CMOS nanoelectronics.

**Acknowledgements.** A.I.K. thanks James McVittie and Carsen Kline for their support and discussions about material deposition. Authors thank Ning Yang for useful discussions on the density functional theory simulation. Authors also thank Taehoon Cheon from Daegu Gyeongbuk Institute of Science and Technology (DGIST) for supporting TEM characterization. L.H. and E.P. acknowledge partial support from the SUPREME JUMP 2.0 center, a Semiconductor Research Corporation and DARPA program. R.K.A.B acknowledges support from the Stanford Graduate Fellowship and NSERC PGS-D Fellowship. M.H. acknowledges partial support from the US Department of Defense through the Grad-uate Fellowship in STEM Diversity Program. Part of this work was performed at the Stanford Nanofabrication Facility (SNF) and Stanford Nano Shared Facilities (SNSF), supported by the National Science Foundation award EECS-2026822. Another part of this work was supported by the National Research Foundation of Korea (NRF) grant funded by the Korean government (MSIT) (RS-2024-00357895).

**Author contributions.** A.I.K. conceived the idea together with L.H. L.H. and A.I.K designed the experiments. A.I.K. formulated the semimetal contact deposition process and optimized it with input from L.H. $WS_2$ growth was performed by Z.Z. and L.H. L.H. performed AFM and Raman characterization, fabricated the devices, and performed electrical measurements with input from A.M. M.H. and L.H. performed and prepared samples for KPFM measurements. TEM and EDS characterization and analysis were performed by H-M.K. I-K.O. and A-R.C with input from A.I.K. R.K.A.B performed the simulations and relevant analysis with input from A.I.K and L.H. L.H., A.I.K., and R.K.A.B. wrote the manuscript together with E.P and A.M. All authors edited the manuscript.

**Competing interests.** A.I.K, L.H., R.K.A.B, E.P and A.M. are listed as co-inventors on a US patent provisional application (patent application number 63/572048) filed by The Board of Trustees of The Leland Stanford Junior University.

**Data and materials availability:** All data needed to evaluate the conclusions in this paper are present in the paper or the supplementary materials.

**Supplementary Information**

Available upon request from the authors and/or when the manuscript is formally published.